\documentclass[10pt,conference]{IEEEtran}

\usepackage{amsmath,amssymb,amsfonts}
\usepackage{bm}
\usepackage{graphicx}
\usepackage{cite}
\usepackage{mathtools}
\usepackage{booktabs}
\usepackage{xcolor}
\usepackage{comment}
\usepackage{acronym}
 \usepackage{amsthm}
\usepackage{tikz}
\usetikzlibrary{arrows.meta,positioning}

\title{Near-Field Physical-Layer Authentication Under Impersonation Attacks}

\author{\IEEEauthorblockN{Hajar El Hassani \IEEEauthorrefmark{1},
Linda Senigagliesi \IEEEauthorrefmark{1}, and Arsenia Chorti \IEEEauthorrefmark{1}\IEEEauthorrefmark{2}}
\IEEEauthorblockA{
\IEEEauthorrefmark{1}ETIS UMR 8051, CY Cergy Paris Universit\'e, ENSEA, CNRS, F-95000, Cergy, France\\
\IEEEauthorrefmark{2}Barkhausen Institut, Dresden gGmbH, Germany\\
Email: hajar.el-hassani@ensea.fr, linda.senigagliesi@ensea.fr,
arsenia.chorti@ensea.fr}
\thanks{The authors have been partially supported by the EC through the Horizon Europe/JU SNS project ROBUST-6G (Grant Agreement no. 101139068), the  Horizon Europe COST Action Project 6G-PHYSEC (CA22168), the CNRS IPAL Project CONNECTING, the ENSEA SRV project RETRO, the CYU TalCyb Chair in Cybersecurity and by the French government under the France 2030 ANR program “PEPR Networks of the Future” (ref. ANR-22-PEFT-0009 HISEC and ANR-22-PEFT-0010 FOUNDS).
}\vspace{-1em}} 

\IEEEoverridecommandlockouts

\begin{document}
\maketitle

\begin{abstract}
This paper studies physical-layer authentication (PLA) in the near-field regime under impersonation attacks. Unlike the far-field case, where the steering vector depends only on the angle of arrival (AoA), in the near field it depends on both angle and distance. We analyze the attack by minimizing the mean-square error (MSE) between the signal received from a legitimate transmitter Alice and the signal generated by an active attacker Eve. For a single-antenna Eve, we derive the optimal scalar precoder and show that, for an inter-element spacing no larger than half a wavelength and under the standard second-order Fresnel approximation, perfect impersonation is possible only if Eve has the same angle and the same distance from Bob as Alice. We then extend the analysis to a multi-antenna Eve and derive the optimal precoding vector. In this case, perfect impersonation is possible only if Alice's steering vector belongs to the subspace spanned by Eve's steering vectors. Simulation results show that, in the near field, a distance difference is sufficient to prevent a successful impersonation attack even when Alice and Eve have the same AoA, and confirm the analytical results.
\end{abstract}


\section{Introduction}

The large-scale deployment of resource-constrained Internet of Things (IoT) devices in beyond-5G and future 6G networks creates important security challenges. In such settings, conventional authentication mechanisms based on upper-layer cryptographic protocols may introduce non-negligible complexity, signaling overhead, and latency. This has motivated the study of lightweight alternatives such as physical-layer authentication (PLA), which exploits device- or channel-dependent physical features of the received signal for transmitter verification. PLA methods are commonly divided into device-based and channel-based approaches \cite{chorti2022context}. The former rely on hardware fingerprints or RF impairments, whereas the latter exploit propagation-related features such as the channel frequency response, channel impulse response, received signal strength, or angle of arrival (AoA).

Among these features, AoA has received increasing attention because it directly captures the spatial signature of the received signal. When the receiver is equipped with an antenna array, the AoA can be estimated from pilot signals using array-processing techniques such as multiple signal classification (MUSIC). In AoA-based PLA, the receiver estimates the AoA of the received signal and compares it with the reference signature of the legitimate transmitter. Existing studies have shown that AoA can be a robust feature against impersonation attacks in digital-array systems \cite{Pham2026,Pham2023GLOBECOM}. In particular, recent works have analyzed the conditions under which a single- or multi-antenna attacker can reproduce the legitimate far-field steering vector \cite{Pham2026}. Related work has also examined the use of AoA for authentication in analog-array multiple-input and multiple-output (MIMO) architectures \cite{Srinivasan2024AnalogAoA}. More generally, AoA has been considered together with other channel-dependent features in channel-based PLA schemes \cite{mitev2023physical,fischer2025systematic}.

The above works are investigated under the far-field assumption, where the transmitter is located beyond the Fraunhofer distance of the array and the received wavefront can be approximated as planar. However, emerging wireless systems increasingly use larger antenna arrays and higher carrier frequencies, such as in millimeter-wave and terahertz communications, large-scale MIMO, and Integrated Sensing and Communication (ISAC) systems. Consequently, under these conditions, many practical communication and sensing scenarios now operate in the near-field region, i.e., the Fresnel region, where the electromagnetic wavefront is spherical rather than planar \cite{an2024near}. This directly affects PLA where, in the far field, the steering vector depends only on AoA, whereas in the near field, it depends on both angle and distance to the transmitter, so the receiver observes a richer spatial signature and the attacker can no longer be characterized by AoA alone.

In this paper, we investigate impersonation attacks in the near-field regime. We analyze the attack by minimizing the mean-square error between the received signatures of a legitimate transmitter and an active adversary. In the far field, exact impersonation by a single-antenna attacker reduces to matching the legitimate AoA \cite{Pham2025AoA}. In the near field, exact impersonation requires matching the full steering signature which leads, under the standard second-order Fresnel approximation, to matching both the angle and the distance. For a multi-antenna attacker, exact impersonation is instead characterized by a span condition on the attacker steering vectors.

To the best of our knowledge, the effect of impersonation attacks on AoA-based PLA has not been explicitly analyzed in the near-field regime. This paper addresses that problem and makes the following contributions:
\begin{itemize}
    \item We formulate a near-field PLA model in which the receiver signature depends on both angle and distance.
     \item For a single-antenna Eve, we derive the optimal scalar precoder that minimizes the MSE and show that, for an inter-element spacing no larger than half a wavelength and under the standard second-order Fresnel approximation, perfect impersonation is possible only if Eve has the same angle and distance from Bob as Alice.
    \item For a multi-antenna Eve, we derive the optimal precoding vector and show that perfect impersonation is possible only if Alice's steering vector lies in the subspace spanned by Eve's steering vectors.
    \item We provide simulation results that show the effect of angle and distance on the minimum MSE, and confirm that in the near field a distance mismatch is sufficient to prevent perfect impersonation even when Alice and Eve have the same AoA.
\end{itemize}

\textit{Organization:} the rest of the paper is organized as follows. Section II introduces the signal model. Section III presents the authentication analysis in the near-field regime. Section IV discusses the numerical results. Finally, Section V concludes the paper.

\section{System Model}

\begin{figure}[t]
\centering
\begin{tikzpicture}[scale=1.0, every node/.style={font=\small}]
    \coordinate (B0) at (0,0);

    \draw[->, thick] (-0.3,0) -- (8.0,0);
    \node[below] at (7.8,-0.05) {$x$};

    \draw[->, thick] (0,-2.3) -- (0,2.3);
    \node[left] at (0,2.25) {$y$};

    \foreach \yy in {-1.75,-1.05,-0.35,0.35,1.05,1.75} {
        \filldraw[black] (0,\yy) circle (2pt);
    }
    \draw[black, dashed, rounded corners] (-0.35,-2.1) rectangle (0.35,2.1);
    \node[above] at (0.3,2.2) {Bob};

    \coordinate (A) at (3.6,1.7);
    \filldraw[blue] (A) circle (2.6pt);
    \node[blue, above] at (3.6,1.9) {Alice};
    \draw[blue, dotted, thick] (B0) -- (A)
        node[midway, above, sloped, blue] {$r_A$};

    \foreach \r in {0.9,1.5,2.1} {
    \draw[blue, dashed, opacity=0.7]
        (A) ++(170:\r) arc[start angle=170,end angle=225,radius=\r];}

    \draw[blue, thick] (1.0,0) arc[start angle=0,end angle=25,radius=1.0];
    \node[blue] at (1.3,0.35) {$\theta_A$};

    \foreach \yy in {0.6,2.2} {
        \filldraw[red] (6.2,\yy) circle (2pt);
    }
    \foreach \yy in {1.1,1.4,1.7} {
        \filldraw[red] (6.2,\yy) circle (0.5pt);
    }
    \draw[red, dashed, rounded corners] (5.85,0.3) rectangle (6.55,2.55);
    \node[above] at (6.2,2.75) {Eve};

    \draw[red, dotted, thick] (B0) -- (6.2,0.6);
    \draw[red, dotted, thick] (B0) -- (6.2,1.4)
        node[midway, above, sloped, red] {$r_{E,l}$};
    \draw[red, dotted, thick] (B0) -- (6.2,2.2);

    \coordinate (E) at (6.2,1.4);
    \foreach \r in {1.0,1.7,2.4} {
    \draw[red, dashed, opacity=0.7]
        (E) ++(170:\r) arc[start angle=170,end angle=225,radius=\r];}

    \draw[red, thick] (2.4,0) arc[start angle=0,end angle=12.7,radius=2.4];
    \node[red] at (2.85,0.48) {$\theta_{E,l}$};

\end{tikzpicture}
\caption{System model for near-field PLA. Bob is equipped with a ULA of \(M\) receive antennas. Alice is equipped with a single antenna located at \((r_A,\theta_A)\). Eve is equipped with a ULA of \(L\) antennas, whose elements are observed under angles \(\theta_{E,l}\), \(l=0,\dots,L-1\).}
\label{fig:system_model}
\end{figure}
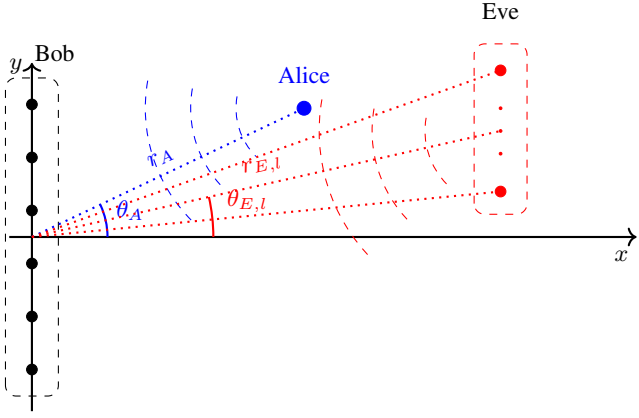

We consider an PLA scenario in which a receiver Bob must decide whether a received signal originates from a legitimate transmitter Alice or from an active attacker Eve. Bob employs a uniform linear array (ULA) of \(M\) receive antennas. Alice is equipped with a single transmit antenna. Eve is equipped either with a single transmit antenna or with \(L\) transmit antennas and applies precoding to alter the signal and try to impersonate Alice.

As depicted in Fig. 1, we adopt a two-dimensional Cartesian coordinate system where Bob's ULA is placed on the \(y\)-axis and centered at the origin. The position of the \(m\)-th receive antenna is denoted by $\mathbf r_m=
\begin{bmatrix}
0 & y_m
\end{bmatrix}^{T},
\ m\in\{0,\dots,M-1\}$, where $y_m$ is the vertical coordinate of the $m$-th antenna. For a centered ULA with inter-element spacing $d \leq \lambda/2$, where $\lambda$ is the wavelength, 
\begin{equation}
y_m=\left(m-\frac{M-1}{2}\right)d.
\label{eq:ym}
\end{equation}

The position of a transmitter is described in polar coordinates by the pair $(r,\theta)$, where $r>0$ denotes its distance from the origin and $\theta\in(-\pi/2,\pi/2)$ is the angle measured from the array broadside.
With this convention, the Cartesian coordinates of the source are $\mathbf p(r,\theta)=
\begin{bmatrix}
r\cos\theta & r\sin\theta
\end{bmatrix}^{T}$.

Alice is located at position $\mathbf p_A=\mathbf p(r_A,\theta_A)$. In the single-antenna adversarial case, Eve is located at $\mathbf p_E=\mathbf p(r_E,\theta_E)$, whereas in the multi-antenna case, Eve's \(L\) transmit antennas are located at $\mathbf p_{E,l}=\mathbf p(r_{E,l},\theta_{E,l}),
\ \forall l\in\{0,\dots,L-1\}.$
\subsection{Near-field steering vector}

For a transmitter located at the position \(\mathbf p(r,\theta)\), the distance to the \(m\)-th receive antenna is
\begin{equation}
d_m(r,\theta)=\|\mathbf p(r,\theta)-\mathbf r_m\|_2,
\qquad m\in\{0,\dots,M-1\}
\label{eq:dm_def}
\end{equation}
where $\|\cdot\|_2$ denotes the Euclidean norm. The distance in \eqref{eq:dm_def} is then computed as
\begin{equation}
d_m(r,\theta)
=
\sqrt{r^2+y_m^2-2ry_m\sin\theta}.
\label{eq:dm}
\end{equation}

Without loss of generality, we choose antenna \(m=0\) as the reference element and define the relative path length
\begin{equation}
\Delta d_m(r,\theta)=d_m(r,\theta)-d_0(r,\theta), \ \forall m
\label{eq:delta_dm}
\end{equation}

Under the narrowband assumption, the spatial signature at Bob is determined by the relative path lengths in \eqref{eq:delta_dm}. The near-field steering vector associated with the transmitter position $(r,\theta)$ is therefore defined as
\begin{equation}
\mathbf a(r,\theta)=
\begin{bmatrix}
1\\
e^{-j\frac{2\pi}{\lambda}\Delta d_1(r,\theta)}\\
e^{-j\frac{2\pi}{\lambda}\Delta d_2(r,\theta)}\\
\vdots\\
e^{-j\frac{2\pi}{\lambda}\Delta d_{M-1}(r,\theta)}
\end{bmatrix}.
\label{eq:nf_steering}
\end{equation}

Unlike the far-field case \cite{Pham2025AoA}, the steering vector in the near field depends on the position, i.e., the joint angle and distance. In what follows, we use the shorthand notation
\[
\mathbf{a_A} \triangleq \mathbf a(r_A,\theta_A),\quad
\mathbf{a_E} \triangleq \mathbf a(r_E,\theta_E),\quad
\mathbf a_{E,l} \triangleq \mathbf a(r_{E,l},\theta_{E,l}),
\]
for Alice, single-antenna Eve, and the \(l\)-th antenna of multi-antenna Eve, respectively.

Let $x\in\mathbb{C}$ denote the signal transmitted during authentication. We assume $\mathbb E[|x|^2]=1$. When Alice transmits, the signal received at Bob is
\begin{equation}
\mathbf y_A=\mathbf a_A\,x+\mathbf n_A,
\label{eq:yA}
\end{equation}
where $\mathbf n_A\sim\mathcal{CN}(\mathbf 0,\sigma_A^2\mathbf I_M)$
is a circularly symmetric complex Gaussian noise vector, $\mathbf I_M$ denotes the $M\times M$ identity matrix and $\sigma_A^2$ is the noise variance per receive antenna.

\subsection{Single-antenna Eve model}

In the single-antenna adversarial case, Eve is located at \(\mathbf p_E\) and applies a complex scalar precoder \(q \in\mathbb C\). The received signal at Bob is
\begin{equation}
\mathbf y_E=\mathbf a_E\,q\,x+\mathbf n_E,
\label{eq:yE_single}
\end{equation}
where $\mathbf n_E\sim\mathcal{CN}(\mathbf 0,\sigma_E^2\mathbf I_M)$.

\subsection{Multi-antenna Eve model}

In the multi-antenna adversarial case, Eve has $L$ transmit antennas located at positions $\mathbf p_{E,l}$, $ \forall l$. The steering vector associated with the $l$-th antenna is $\mathbf a_{E,l}$. By stacking these vectors into a matrix form, we obtain
\begin{equation}
\mathbf A_E=
\begin{bmatrix}
\mathbf a_{E,0} & \mathbf a_{E,1} & \cdots & \mathbf a_{E,L-1}
\end{bmatrix}
\in\mathbb C^{M\times L}.
\label{eq:AE}
\end{equation}

Eve applies the precoding vector
\begin{equation}
\mathbf q=
\begin{bmatrix}
q_0 & q_1 & \cdots & q_{L-1}
\end{bmatrix}^{T}
\in\mathbb C^{L\times 1}.
\label{eq:q}
\end{equation}
Hence, the signal received at Bob is
\begin{equation}
\mathbf y_E=\mathbf A_E\mathbf q\,x+\mathbf n_E.
\label{eq:yE_multi}
\end{equation}

As shown in \eqref{eq:yE_multi}, in the multi-antenna case, Eve transmits a linear combination of the steering vectors associated with each of her antennas position.

In the next section, we determine under which conditions Eve can choose the precoder so that the signal received at Bob becomes indistinguishable from Alice's signal in terms of its spatial signature.

\section{Near-Field Authentication Analysis}

We now study the conditions under which an impersonation attack can be successful in the near-field regime. Specifically, we examine whether Eve can alter the transmitted signal through precoding so that the spatial signature received at Bob becomes indistinguishable from Alice's. We consider a worst-case attacker with perfect knowledge of Alice’s and Bob’s positions and array geometry, as well as ideal precoding capabilities. Practical limitations, such as imperfect knowledge, synchronization errors, and hardware constraints, would make successful impersonation more difficult for Eve. To this end, we formulate the problem by minimizing the mean-square error (MSE) between the signal received from Alice and the signal received from Eve. We first consider the case in which Eve is equipped with a single antenna, and then extend the analysis to the multi-antenna case.

\subsection{Single-antenna Eve}

In the single-antenna adversarial case, the MSE between the received signals corresponding to Alice and Eve is defined as
\begin{equation}
\zeta(q)=\mathbb E\!\left[\|\mathbf y_A-\mathbf y_E\|^2\right].
\label{eq:mse_def}
\end{equation}

Using \eqref{eq:yA} and \eqref{eq:yE_single}, we obtain
\begin{equation}
\mathbf y_A-\mathbf y_E
=
\left(\mathbf a_A-q\mathbf a_E\right)x + (\mathbf n_A-\mathbf n_E).
\label{eq:y_difference}
\end{equation}
Substituting \eqref{eq:y_difference} into \eqref{eq:mse_def} yields
\begin{equation}
\zeta(q)
=
\mathbb E\!\left[
\left\|
\left(\mathbf a_A-q\mathbf a_E\right)x + (\mathbf n_A-\mathbf n_E)
\right\|^2
\right].
\label{eq:mse_expand_start}
\end{equation}

Given that $\mathbb E[|x|^2]=1$,
and that the noise terms are zero-mean and independent of the signal, the cross terms vanish after expectation. Therefore,
\begin{equation}
\zeta(q)
=
\left\|
\mathbf a_A-q\mathbf a_E
\right\|^2
+
\frac{1}{\gamma_A}
+
\frac{1}{\gamma_{E}},
\label{eq:mse_decomp}
\end{equation}
where \(\gamma_A\) and \(\gamma_E\) denote the signal-to-noise ratios (SNRs) of the legitimate and adversarial links, respectively. The impersonation problem can thus be written as
\begin{equation}
q^\star
=
\arg\min_{q\in\mathbb C}\zeta(q).
\label{eq:opt_problem_mse}
\end{equation}
Since the noise terms in \eqref{eq:mse_decomp} do not depend on \(q\), \eqref{eq:opt_problem_mse} is equivalent to
\begin{equation}
q^\star
=
\arg\min_{q\in\mathbb C}
\left\|
\mathbf a_A-q\mathbf a_E
\right\|^2.
\label{eq:opt_problem_reduced}
\end{equation}

Expanding the quadratic term gives
\begin{align}
\left\|\mathbf a_A-q\mathbf a_E\right\|^2
&=
(\mathbf a_A-q\mathbf a_E)^H(\mathbf a_A-q\mathbf a_E)
\nonumber\\
&=
\mathbf a_A^H\mathbf a_A
-
q\,\mathbf a_A^H\mathbf a_E
-
q^*\,\mathbf a_E^H\mathbf a_A
+
|q|^2\,\mathbf a_E^H\mathbf a_E.
\label{eq:mse_expanded}
\end{align}

Since $\zeta(q)$ is a real-valued function, we use Wirtinger calculus and differentiate w.r.t $q^*$. From \eqref{eq:mse_expanded}, we obtain
\begin{equation}
\frac{\partial \zeta(q)}{\partial q^*}
=
-\mathbf a_E^H\mathbf a_A
+
q\,\mathbf a_E^H\mathbf a_E.
\label{eq:df_dqstar}
\end{equation}
Setting \eqref{eq:df_dqstar} to zero yields the unique global minimizer
\begin{equation}
q^\star
=
\frac{\mathbf a_E^H\mathbf a_A}{\mathbf a_E^H\mathbf a_E}.
\label{eq:q_opt_single}
\end{equation}

Equation \eqref{eq:q_opt_single} gives the scalar precoder that minimizes the mismatch between Alice's and Eve's steering vectors. Exact impersonation is possible only if the minimum mismatch is zero, i.e., $\left\|\mathbf a_A-q^\star\mathbf a_E\right\|^2=0$, which is equivalent to
\begin{equation}
\mathbf a_A=q^\star\mathbf a_E.
\label{eq:exact_impersonation}
\end{equation}

The steering-vector expression in \eqref{eq:nf_steering} implies $q^\star=1$. Hence, exact impersonation is possible only if
\begin{equation}
\mathbf a_A=\mathbf a_E.
\label{eq:a_eq_ahat}
\end{equation}

We now determine under which condition \eqref{eq:a_eq_ahat} holds. Since the steering vectors are built from the relative path lengths, equality of steering vectors implies
\begin{equation}
\Delta d_m(r_A,\theta_A)-\Delta d_m(r_E,\theta_E) = k_m\lambda, \ k_m\in\mathbb{Z}, \   \forall m
\label{eq:delta_equal}
\end{equation}

Since $\Delta d_0=0$, we have $k_0=0$. Subtracting \eqref{eq:delta_equal} for two consecutive antenna elements 
\begin{equation}
(d_{m,A}-d_{m-1,A})
-
(d_{m,E}-d_{m-1,E})=(k_m-k_{m-1})\lambda
\label{consecutive_antenna}
\end{equation}

Moreover, for any transmitter $X\in\{A,E\}$, it can be shown that,
for $\theta_X\in(-\pi/2,\pi/2)$
\begin{equation}
    |d_{m,X}-d_{m-1,X}|<d
\end{equation}

Hence, by taking the absolute value of \eqref{consecutive_antenna} and applying the triangle inequality, and given that $d\leq\lambda/2$, we have
\begin{equation}
|(k_m-k_{m-1})|\lambda
<
2d
\leq \lambda
\end{equation}
which implies $k_m=k_{m-1}$. Since $k_0=0$, it follows recursively that $k_m=0, \forall m$, and therefore
\begin{equation}
\Delta d_m(r_A,\theta_A)
=
\Delta d_m(r_E,\theta_E),
\quad \forall m\in\{0,\ldots,M-1\}
\end{equation}

The distance in \eqref{eq:dm} is written as
\begin{equation}
d_m(r,\theta)=r\sqrt{1+\varepsilon_m},
\label{eq:dm_eps}
\end{equation}
where $\varepsilon_m  = y_m^2/r^2 - 2 y_m \sin\theta / r$.
When the array aperture is small compared with the transmitter distance, we have $|\varepsilon_m|\ll 1$, and the second-order Taylor expansion gives \cite{ebadi2025near,qu2024near}
\begin{equation}
\sqrt{1+\varepsilon_m}
\approx
1+\frac{\varepsilon_m}{2}-\frac{\varepsilon_m^2}{8}.
\end{equation}
Substituting $\varepsilon_m$ and retaining terms up to second order yields
\begin{equation}
d_m(r,\theta)
\approx
r-y_m\sin\theta+\frac{y_m^2}{2r}\cos^2\theta,
\label{eq:fresnel_dm}
\end{equation}
which is called the Fresnel approximation \cite{Fresnel_approx} and was proved in \cite{taylor_approx_proof} that the second-order expansion is accurate enough in the near-field region. This approximation is valid in the radiative near-field region considered in this work, but may become less accurate in the extreme near field or in environments with strong scattering. Accordingly, the relative path length in \eqref{eq:delta_dm} becomes
\begin{equation}
\Delta d_m(r,\theta)
\approx
-(y_m-y_0)\sin\theta
+
\frac{y_m^2-y_0^2}{2r}\cos^2\theta,
\ \forall m
\label{eq:fresnel_delta_dm}
\end{equation}

Applying \eqref{eq:fresnel_delta_dm} to \eqref{eq:delta_equal} yields
\begin{equation}
\begin{split}
(y_m-y_0)\Biggl[
&(\sin\theta_A-\sin\theta_E) \\
&\quad + \frac{y_m+y_0}{2}
\left(
\frac{\cos^2\theta_E}{r_E}
-
\frac{\cos^2\theta_A}{r_A}
\right)
\Biggr]=0,
\ \forall m
\end{split}
\label{eq:lin_quad_eq}
\end{equation}
For the reference antenna \(m=0\), the factor \((y_m-y_0)\) is zero, so \eqref{eq:lin_quad_eq} is directly satisfied. For any antenna with \(y_m\neq y_0\), we must have
\begin{equation}
(\sin\theta_A-\sin\theta_E)
+
\frac{y_m+y_0}{2}
\left(
\frac{\cos^2\theta_E}{r_E}
-
\frac{\cos^2\theta_A}{r_A}
\right)
=0.
\label{eq:condition}
\end{equation}

Now consider two non-reference antennas \(m_p\) and \(m_q\), with \(m_p\neq m_q\), so that \(y_{m_p}\neq y_{m_q}\). Writing \eqref{eq:condition} for \(m_p\) and \(m_q\) gives
\begin{equation}
(\sin\theta_A-\sin\theta_E)
+
\frac{y_{m_p}+y_0}{2}
\left(
\frac{\cos^2\theta_E}{r_E}
-
\frac{\cos^2\theta_A}{r_A}
\right)
=0,
\label{eq:mp}
\end{equation}
\begin{equation}
(\sin\theta_A-\sin\theta_E)
+
\frac{y_{m_q}+y_0}{2}
\left(
\frac{\cos^2\theta_E}{r_E}
-
\frac{\cos^2\theta_A}{r_A}
\right)
=0,
\label{eq:mq}
\end{equation}
Subtracting \eqref{eq:mq} from \eqref{eq:mp}, we obtain
\begin{equation}
\left(
\frac{\cos^2\theta_E}{r_E}
-
\frac{\cos^2\theta_A}{r_A}
\right)
(y_{m_p}-y_{m_q})=0.
\end{equation}
Since \(y_{m_p}\neq y_{m_q}\), it follows that
\begin{equation}
\frac{\cos^2\theta_E}{r_E}
=
\frac{\cos^2\theta_A}{r_A}
\label{eq:range_relation}.
\end{equation}
Substituting \eqref{eq:range_relation} back into either \eqref{eq:mp} or \eqref{eq:mq} yields
\begin{equation}
\sin\theta_A=\sin\theta_E.
\label{eq:sin_relation}
\end{equation}

Since \(\theta_A,\theta_E\in(-\pi/2,\pi/2)\), the equality in \eqref{eq:sin_relation} implies
\begin{equation}
\theta_E=\theta_A.
\label{eq:theta_equal}
\end{equation}
Substituting \eqref{eq:theta_equal} in \eqref{eq:range_relation} yields
\begin{equation}
r_E=r_A.
\label{eq:r_equal}
\end{equation}

Therefore, under the second-order Fresnel approximation, exact impersonation in the single-antenna case is possible only if Eve is at the same position, i.e., the same angle and distance from Bob.

\subsection{Multi-antenna Eve}

We now consider the case in which Eve is equipped with \(L\) transmit antennas. Using \eqref{eq:yA} and \eqref{eq:yE_multi} in \eqref{eq:mse_def}, and computing the expectation as in the single-antenna case, the MSE can be written as
\begin{equation}
\zeta(\mathbf q)
=
\left\|
\mathbf a_A-\mathbf A_E\mathbf q
\right\|^2
+
\frac{1}{\gamma_A}
+
\frac{1}{\gamma_E}.
\label{eq:MSE_multi}
\end{equation}

The corresponding impersonation problem is
\begin{equation}
\mathbf q^\star
=
\arg\min_{\mathbf q\in\mathbb C^{L\times 1}}
\left\|
\mathbf a_A-\mathbf A_E\mathbf q
\right\|^2.
\label{eq:opt_problem_multi_reduced}
\end{equation}

Expanding the quadratic term gives
\begin{align}
\left\|
\mathbf a_A-\mathbf A_E\mathbf q
\right\|^2
&=
(\mathbf a_A-\mathbf A_E\mathbf q)^H
(\mathbf a_A-\mathbf A_E\mathbf q)
\nonumber\\
&=
\mathbf a_A^H\mathbf a_A
-
\mathbf a_A^H\mathbf A_E\mathbf q
-
\mathbf q^H\mathbf A_E^H\mathbf a_A \nonumber \\ 
& \qquad+
\mathbf q^H\mathbf A_E^H\mathbf A_E\mathbf q.
\label{eq:mse_multi_expanded}
\end{align}

Similarly, we differentiate \eqref{eq:mse_multi_expanded} w.r.t \(\mathbf q^*\) and set it to zero to obtain 
\begin{equation}
\mathbf A_E^H\mathbf A_E\,\mathbf q^\star
=
\mathbf A_E^H\mathbf a_A.
\label{eq:q_opt_multi}
\end{equation}

Hence, \eqref{eq:q_opt_multi} has the unique minimum-norm solution
\begin{equation}
\mathbf q^\star
=
\mathbf A_E^\dagger \mathbf a_A
\label{eq:q_opt_multi_pinv},
\end{equation}
where $\mathbf A_E^\dagger$ denotes the Moore--Penrose pseudoinverse.

Equation \eqref{eq:q_opt_multi_pinv} gives the precoding vector that minimizes the mismatch between Alice's steering vector and the linear combination of Eve's steering vectors. Exact impersonation is possible only if the mismatch at the optimum is zero, i.e., $\left\|
\mathbf a_A-\mathbf A_E\mathbf q^\star
\right\|^2=0$, which is equivalent to
\begin{equation}
\mathbf a_A=\mathbf A_E\mathbf q^\star
=
\sum_{l=0}^{L-1} q_l^\star \mathbf a_{E,l}.
\label{eq:nf_multi_cond}
\end{equation}

Hence, exact impersonation is possible only if Alice's steering vector belongs to the subspace spanned by Eve's steering vectors as
\begin{equation}
\mathbf a_A \in \mathrm{span}\left\{\mathbf a_{E,0},\mathbf a_{E,1},\dots,\mathbf a_{E,L-1}\right\}.
\label{eq:span_condition}
\end{equation}

We note that Eve cannot alter the steering vectors associated with her antenna positions, since these are fixed by geometry. Hence, the only degree of freedom available to Eve is the linear combination produced through $\mathbf q$. The optimal precoding vector is therefore the one that projects Alice's steering vector onto the subspace generated by Eve's steering vectors. If \(\mathbf a_A\) lies outside this subspace, then a nonzero mismatch will remain and perfect impersonation is impossible.

\section{Simulation Results}

This section presents numerical results that validate the analytical derivations developed in the previous sections.

\subsection{Simulation setup}

We consider a carrier frequency of $f = 2.18$~GHz, corresponding to a wavelength of $\lambda \approx 0.138$~m. The legitimate receiver Bob is equipped with a ULA of $M = 16$. The far field starts at the Fraunhofer distance $d_F = 2\frac{D^2}{\lambda}$ from the transmitter \cite{Bjornson2021, Selvan2017}, where $D$ represents the aperture of the array. In our simulations, this parameter corresponds to $D = (M-1)\,d \approx 1.034$~m, where $d=\lambda/2$. Hence, under the selected frequency, the distance limit between the near and far field is set as $d_F \approx 15.5$~m.

\subsection{Simulation figures}

\begin{figure}
    \centering
    \includegraphics[width=0.45\textwidth]{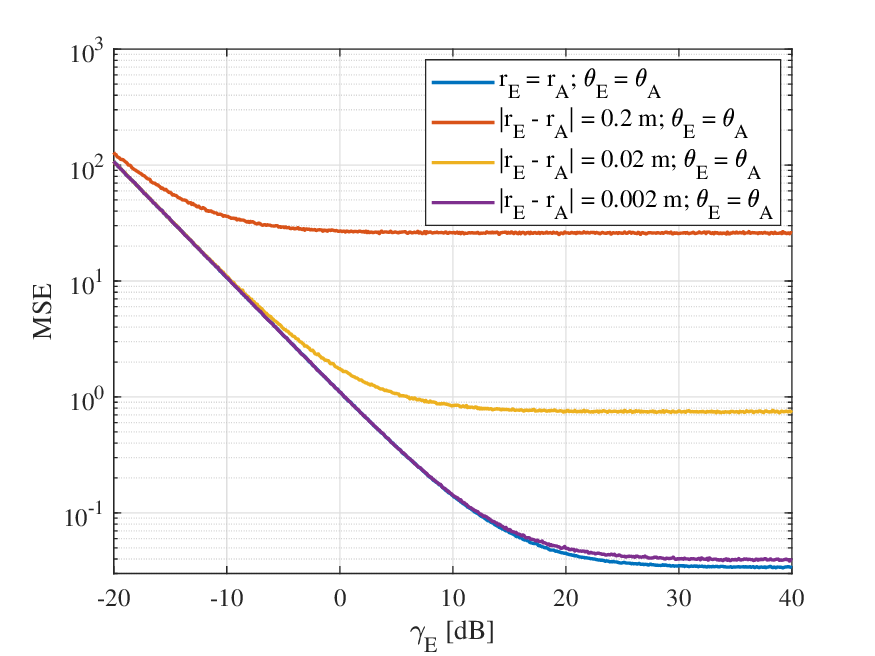}
    \caption{MSE vs SNR. Alice and Eve with a single antenna. $\theta_A = \theta_E$, $\gamma_A$ = $15$~dB. }
    \label{fig:mse_snr}
\end{figure}

First, we show the impact of the SNR related to the active attacker Eve on the MSE, defined in \eqref{eq:mse_def}, in the single-antenna case. In Fig. \ref{fig:mse_snr}, the SNR at Alice is fixed at $\gamma_A = 15$~dB, while Eve's SNR is varying between $-20$ and $40$~dB. Both Alice and Eve are assumed to have the same AoA, i.e., $\theta_A = \theta_E = 0.4$ rad, representing a challenging authentication scenario in which the two sources are angularly co-located as seen from Bob's array. The MSE is obtained by averaging over $ 10^4$ independent Monte Carlo realizations. 
The figure shows the MSE as a function of the adversarial SNR \(\gamma_E\) for different values of the distance mismatch \(|r_E-r_A|\). First, for all curves, the MSE decreases as \(\gamma_E\) increases, since the term \(1/\gamma_E\) becomes smaller. Second, when \(r_E=r_A\), the mismatch term is zero and the MSE reduces to \( (1/\gamma_A+1/\gamma_E) \). By contrast, when \(r_E\neq r_A\), the mismatch term remains nonzero even for large \(\gamma_E\), and the MSE converges to a larger value. This value increases with the distance mismatch \(|r_E-r_A|\). Hence, in the near field, exact impersonation is possible only if the attacker matches both the angle and the distance relative to Bob.

\begin{figure}
    \centering
    \includegraphics[width=0.45\textwidth]{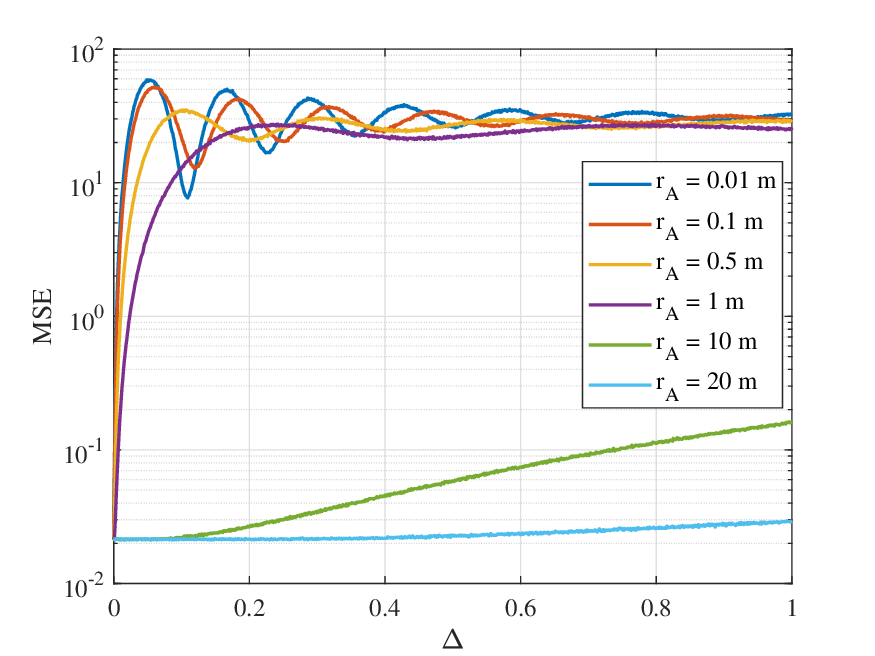}
    \caption{MSE vs Alice-Eve distance. $\Delta = |r_A - r_E|$, with $\gamma_A = \gamma_E = 20$~dB. }
    \label{fig:mse_re}
\end{figure}

In Fig. \ref{fig:mse_re} we further investigate the role of the distance mismatch between Alice and Eve, denoted as $\Delta = |r_A - r_E|$, for different values of Alice's distance \(r_A\), under the assumption \(\theta_A=\theta_E\) and \(\gamma_A=\gamma_E=20\) dB. As expected from the analysis, the MSE reaches its minimum value only at \(\Delta=0\), i.e., when Alice and Eve are co-located. When \(\Delta\neq 0\), the MSE becomes nonzero, and its variation with \(\Delta\) depends strongly on \(r_A\). For small and moderate values of \(r_A\), the MSE changes significantly with \(\Delta\), which shows a strong dependence of the steering vector on the distance. For larger values of \(r_A\), the curves become flatter and remain closer to the minimum value. In particular, low MSE values are observed for distances on the order of \(10\) m, which is closer to the Fraunhofer distance, where the dependence of the steering vector on the distance becomes weaker. Therefore, the results confirm that, in the near field, a distance mismatch is sufficient to prevent exact impersonation even when Alice and Eve have the same AoA.

\begin{figure}
    \centering
    \includegraphics[width=0.45\textwidth]{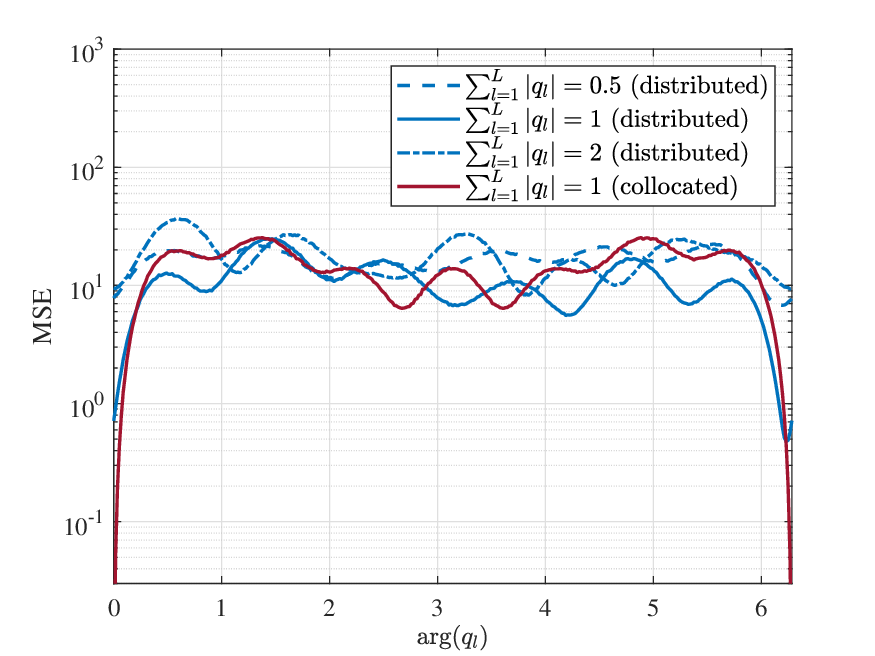}
    \caption{Effect of precoding $\bm{q}$ on MSE. Alice with a single antenna, Eve with $L=8$ antennas. $\theta_A = 0.4$ rad, $r_A = 1$ m. $\gamma_A = \gamma_E = 40$ dB. }
    \label{fig:precoding}
\end{figure}

Finally, we show the effect of the precoding vector \(\mathbf q\) in the multi-antenna Eve case. In Fig.~\ref{fig:precoding}, Alice is located at \((r_A,\theta_A)=(1~\mathrm{m},0.4~\mathrm{rad})\), while Eve is equipped with \(L=8\) antennas. Two scenarios are considered. In the collocated case, all Eve antennas are placed at the same position as Alice, so that \(\mathbf a_{E,l}=\mathbf a_A, \forall l\). This is a special case of the condition in \eqref{eq:span_condition}. In this case, \(\mathbf A_E=\mathbf a_A\mathbf 1^T\), and the exact impersonation condition is satisfied if \(\sum_{l=0}^{L-1} q_l=1\).  In the distributed case, Eve's antennas are placed at different positions in a small region around Alice, where the condition in \eqref{eq:span_condition} is not satisfied. For simplicity, we consider that all entries of the precoding vector share the same phase $\arg(q_l)$. The figure shows the MSE as a function of the common phase for different values of \(\sum_{l=1}^{L}|q_l|\).

The collocated case achieves the lowest MSE among all considered settings. In particular, when $\arg(q_l)=0$, the exact impersonation condition is satisfied and the MSE reduces to the minimum value \((1/\gamma_A+1/\gamma_E)\), whereas the distributed case gives a significantly larger MSE, as expected from the analysis. The MSE remains significantly larger in both scenarios for different values of $\arg(q_l)$ and \(\sum_{l=1}^{L}|q_l|\). This shows that, in the multi-antenna case, exact impersonation is possible only if the precoding vector of Eve is chosen such that Alice's steering vector belongs to the subspace spanned by Eve's steering vectors, which is consistent with the analysis.

\section{Conclusion}
This paper investigates impersonation attacks on physical-layer authentication in the near field where the steering vector depends on both angle and distance.
Using an MSE-based formulation, we first analyzed the single-antenna Eve case. We derived the optimal scalar precoder and showed that exact impersonation requires Eve to reproduce the same steering vector as Alice. For an inter-element spacing no larger than half a wavelength and under the standard second-order Fresnel approximation, this implies that Eve must have the same angle and the same distance relative to Bob. We then considered the multi-antenna Eve case where exact impersonation is possible only if Alice's steering vector belongs to the subspace spanned by Eve's steering vectors.
The results show that the conditions for successful impersonation are stricter in the near field than in the far field, where exact impersonation cannot be achieved by matching the AoA alone. Future work will investigate propagation conditions where the second-order Fresnel approximation becomes inaccurate, as well as extensions to three-dimensional and MIMO settings.

\section*{Acknowledgment}
The authors would like to thank Alessandro Soranzo from University of Trieste for his valuable comments that helped improve the analysis.

\bibliographystyle{IEEEtran}
\bibliography{references}

@article{ebadi2025near,
  title={Near-field localization with antenna arrays in the presence of direction-dependent mutual coupling},
  author={Ebadi, Zohreh and Molaei, Amir Masoud and Alexandropoulos, George C and Abbasi, Muhammad Ali Babar and Cotton, Simon and Tukmanov, Anvar and Yurduseven, Okan},
  journal={IEEE Transactions on Vehicular Technology},
  year={2025},
  publisher={IEEE}
}

@article{Pham2025AoA,
  author    = {Thuy M. Pham and Linda Senigagliesi and Marco Baldi and Rafael F. Schaefer and Gerhard P. Fettweis and Arsenia Chorti},
  title     = {Leveraging Angle of Arrival Estimation against Impersonation Attacks in Physical Layer Authentication},
  journal   = {CoRR},
  volume    = {abs/2503.11508},
  year      = {2025}
}

@ARTICLE{Pham2026,
  author={Pham, Thuy M. and Senigagliesi, Linda and Baldi, Marco and Schaefer, Rafael F. and Fettweis, Gerhard P. and Chorti, Arsenia},
  journal={IEEE Transactions on Information Forensics and Security}, 
  title={Leveraging Angle of Arrival Estimation Against Impersonation Attacks in Physical Layer Authentication}, 
  year={2026},
  volume={21},
  number={},
  pages={3226-3239},
  doi={10.1109/TIFS.2026.3675885}}

@inproceedings{Pham2023GLOBECOM,
  author    = {Thuy M. Pham and Linda Senigagliesi and Marco Baldi and Gerhard P. Fettweis and Arsenia Chorti},
  title     = {Machine Learning-Based Robust Physical Layer Authentication Using Angle of Arrival Estimation},
  booktitle = {Proc. IEEE Global Communications Conference (GLOBECOM)},
  year      = {2023},
  pages     = {13--18}
}

@inproceedings{Srinivasan2024AnalogAoA,
  author    = {M. Srinivasan and L. Senigagliesi and H. Chen and A. Chorti and M. Baldi and H. Wymeersch},
  title     = {{AoA}-Based Physical Layer Authentication in Analog Arrays Under Impersonation Attacks},
  booktitle = {Proc. IEEE SPAWC},
  year      = {2024},
  pages     = {496--500}
}

@article{qu2024near,
  title={Near-field integrated sensing and communication: Performance analysis and beamforming design},
  author={Qu, Kaiqian and Guo, Shuaishuai and Saeed, Nasir and Ye, Jia},
  journal={IEEE Open Journal of the Communications Society},
  volume={5},
  pages={6353--6366},
  year={2024},
  publisher={IEEE}
}

@article{taylor_approx_proof,
  title={Fraunhofer and Fresnel distances: Unified derivation for aperture antennas},
  author={Selvan, Krishnasamy T and Janaswamy, Ramakrishna},
  journal={IEEE antennas and propagation magazine},
  volume={59},
  number={4},
  pages={12--15},
  year={2017},
  publisher={IEEE}
}

@article{Fresnel_approx,
  title={Localization of signals in the near-field of an antenna array},
  author={Friedlander, Benjamin},
  journal={IEEE Transactions on Signal Processing},
  volume={67},
  number={15},
  pages={3885--3893},
  year={2019},
  publisher={IEEE}
}

@article{mitev2023physical,
  title={Physical layer security—From theory to practice},
  author={Mitev, Miroslav and Pham, Thuy M and Chorti, Arsenia and Barreto, Andr{\'e} Noll and Fettweis, Gerhard},
  journal={IEEE BITS the Information Theory Magazine},
  volume={3},
  number={2},
  pages={67--79},
  year={2023},
  publisher={IEEE}
}

@article{fischer2025systematic,
  title={A systematic survey and comparative analysis of angular-based indoor localization and positioning technologies},
  author={Fischer, Georg KJ and Schaechtle, Thomas and Gabbrielli, Andrea and Bordoy, Joan and H{\"a}ring, Ivo and H{\"o}flinger, Fabian and Rupitsch, Stefan J},
  journal={IEEE Communications Surveys \& Tutorials},
  year={2025},
  publisher={IEEE}
}

@article{chorti2022context,
  title={Context-aware security for {6G} wireless: The role of physical layer security},
  author={Chorti, Arsenia and Barreto, Andr{\'e} Noll and K{\"o}psell, Stefan and Zoli, Marco and Chafii, Marwa and Sehier, Philippe and Fettweis, Gerhard and Poor, H Vincent},
  journal={IEEE Communications Standards Magazine},
  volume={6},
  number={1},
  pages={102--108},
  year={2022},
  publisher={IEEE}
}

@INPROCEEDINGS{Bjornson2021,
  author={Bj{\"o}rnson, Emil and Demir, {\"O}zlem Tu{\u{g}}fe and Sanguinetti, Luca},
  booktitle={2021 55th Asilomar Conference on Signals, Systems, and Computers}, 
  title={A Primer on Near-Field Beamforming for Arrays and Reconfigurable Intelligent Surfaces}, 
  year={2021},
  volume={},
  number={},
  pages={105-112},
  doi={10.1109/IEEECONF53345.2021.9723331}
}

@ARTICLE{Selvan2017,
  author={Selvan, Krishnasamy T. and Janaswamy, Ramakrishna},
  journal={IEEE Antennas and Propagation Magazine}, 
  title={Fraunhofer and {Fresnel} Distances: Unified derivation for aperture antennas}, 
  year={2017},
  volume={59},
  number={4},
  pages={12-15},
  doi={10.1109/MAP.2017.2706648}
  }

@article{an2024near,
  title={Near-field communications: Research advances, potential, and challenges},
  author={An, Jiancheng and Yuen, Chau and Dai, Linglong and Di Renzo, Marco and Debbah, M{\'e}rouane and Hanzo, Lajos},
  journal={IEEE Wireless Communications},
  volume={31},
  number={3},
  pages={100--107},
  year={2024},
  publisher={IEEE}
}

\end{document}